\documentclass[pra,twocolumn]{revtex4}%
\usepackage{amsfonts}
\usepackage{amsmath}
\usepackage{amssymb}
\usepackage{graphicx}%
\begin{document}
\title{Perturbation theory in a pure exchange non-equilibrium economy }
\author{Samuel E. V\'azquez }
\affiliation{Perimeter Institute for Theoretical Physics, Waterloo N2L 2Y5, ON Canada}
\author{Simone Severini}
\affiliation{Institute for Quantum Computing and Department of Combinatorics \&
Optimization University of Waterloo, Waterloo N2L 3G1, ON Canada}

\begin{abstract}
We develop a formalism to study linearized perturbations around the
equilibria of a pure exchange economy. With the use of mean field theory
techniques, we derive equations for the flow of products in an economy driven
by heterogeneous preferences and probabilistic interaction between agents. We
are able to show that if the economic agents have static preferences, which
are also homogeneous in any of the steady states, the final wealth
distribution is independent of the dynamics of the non-equilibrium theory. In
particular, it is completely determined in terms of the initial conditions, and it is independent of the
probability, and the network of interaction between agents. We show that the
main effect of the network is to determine the relaxation time via the usual
eigenvalue gap as in random walks on graphs.

\end{abstract}
\maketitle

\section{Introduction}

There is a growing consensus for the need of a non-equilibrium theory of
economics \cite{bfl, fg, sf}. From a purely
theoretical perspective, one would like to understand how the economy chooses
one of the multitude of possible equilibria. An important and perhaps more
practical question, is to determine which of the equilibrium states are stable.
In other words, given a small perturbation away from such a state, does the
system relaxes back to the same equilibrium, or does it settles to a
completely different state? This is similar to the so-called \emph{landscape
problem} found in some areas of physics such as string theory, frustrated
magnets and protein folding.

Constructing a full non-linear economic theory out of equilibrium presents many
problems. Is price a meaningful concept out of equilibrium? How do
we model the interactions of many heterogeneous agents? \emph{Etc.} Usually
these questions are tackled by computer simulations using \emph{agent-based
models} \cite{b}.

In this paper we consider the simpler case of a pure exchange economy with no
production. We develop an analytic formalism to study linearized perturbations
around any equilibrium state. Our approach is probabilistic, and we make heavy
use of mean field theory techniques. Before setting up the perturbation
theory, we study the landscape of equilibria for the pure exchange economy.
Since the dynamics of the economy should not depend on the units used to
measure the different products, there is a kind of ``gauge"
symmetry in the problem \cite{s}. We show that this symmetry induces an
equivalence relation in the landscape of equilibria. In fact, in the limit of
many agents, we show that the set of equivalence classes of economic
equilibria is in one-to-one correspondence with the space of wealth distributions.

One of the main questions that we ask is: what is the importance of the trading
network topology in determining the final state of the economy? For a related study,  see  \cite{w} and references therein. We
find that, under some more restrictive assumptions on the nature of the
possible equilibrium states, the final state is completely determined in terms
of the initial conditions of the linearized perturbation. In particular, it is
independent of the details of the non-equilibrium dynamics and trading network
structure. We find that the main role of the network topology is to determine
the relaxation time. In our approach, prices are emergent and
describe the relative flow of products between different agents.

The structure of the paper is as follows. In Section II, we study the
landscape of equilibria of the pure exchange economy. In Section III, we describe the probabilistic rules that drive the
dynamics of the system, and derive mean field theory evolution equations for the linearized perturbations. In Section IV, we state and prove our main result regarding the universality of the final wealth distribution.
Section V contains two examples with specific indices of satisfaction. In the
first example, we deal with homogeneous static preferences. We show the corresponding relaxation to equilibrium, and the relation between the relaxation time and the network topology.  In the second example, we study
heterogeneous dynamic preferences. More precisely, we take agents that update their preferences as they trade. Again, we show how the network topology affects the relaxation time to equilibrium. Conclusions are drawn in Section VI.

\section{The landscape of pure exchange equilibria}

Our system includes a set of \emph{agents} $\mathcal{A}=\{\alpha
:\alpha=1,2,\ldots,m\}$ and a set of \emph{products} $\mathcal{P=\{}%
i:i=1,2,\ldots,p\}$. The amount of product $i$ owned by agent $\alpha$ is
denoted by $n_{\alpha}^{i}$. We work with continuous variables, but the
results can also be recasted in a discrete setting. We shall define an
\emph{index of satisfaction} $\Omega_{\alpha}$ specifying the preferences for
each agent $\alpha$. We assume the following basic properties: \emph{(i)
}$\partial_{i}\Omega_{\alpha}>0$, \emph{(ii) }$\partial_{i}^{2}\Omega_{\alpha
}<0$, and \emph{(iii) }$\lim_{n_{\alpha}^{i}\rightarrow\infty}\partial
_{i}\Omega_{\alpha}=\infty$. Here $\partial_{i}$ is a shorthand notation for
the derivative $\partial/\partial n_{\alpha}^{i}$. Since we are considering a
pure exchange economy, it follows that $\frac{d}{dt}\sum_{\alpha}n_{\alpha
}^{i}=0$, where $t$ denotes time. When agents have time changing preferences,
we have $\partial\Omega_{\alpha}/\partial t\neq0$. Since the amount of each
product can be measured in arbitrary units, the dynamics of the economy should
be invariant under a transformation%
\begin{equation}%
\begin{tabular}
[c]{lll}%
$n_{\alpha}^{i}\mapsto\phi^{i}n_{\alpha}^{i},$ & where & $\phi^{i}%
\in\mathbb{R}^{+}.$%
\end{tabular}
\ \ \ \ \ \label{gauge}%
\end{equation}
This defines an equivalence class of all $n_{\alpha}^{i}$ and $\widetilde
{n}_{\alpha}^{i}$ (even at different times) such that $n_{\alpha}^{i}=\phi
^{i}\widetilde{n}_{\alpha}^{i}$. Additionally, we assume that $\partial
_{i}\Omega_{\alpha}\rightarrow(\phi^{i})^{-1}\partial_{i}\Omega_{\alpha}$. Let
us denote by $M_{j}^{i}$ the \emph{exchange rate} of the products $i$ and $j$.
Thus, $M_{j}^{i}\mapsto\phi^{i}(\phi^{j})^{-1}M_{j}^{i}$. \emph{Equilibrium}
(or \emph{steady state})\ in a pure exchange economy is a set of inventories
$\{\bar{n}_{\alpha}^{i}:\alpha\in\mathcal{A},i\in\mathcal{P}\}$ and exchange
rates $\{\bar{M}_{j}^{i}:i,j\in\mathcal{P}\}$ that satisfy the maximization
conditions
\[
\left.  \frac{\partial}{\partial\Delta n^{i}}\Omega_{\alpha}(\bar{n}_{\alpha
}^{i}+\Delta n^{i},\bar{n}_{\alpha}^{j}-\bar{M}_{i}^{j}\Delta n^{i}%
,\ldots)\right\vert _{\Delta n^{i}=0}=0,
\]
leading to
\begin{equation}
\left.  (\partial_{i}\Omega_{\alpha}-\bar{M}_{i}^{j}\partial_{j}\Omega
_{\alpha})\right\vert _{\bar{n}_{\alpha}}=0\label{equilibrium}%
\end{equation}
The solutions of Eq. (\ref{equilibrium}) form equivalence classes under the
transformation $\phi^{i}$. Eq. (\ref{equilibrium}) implies a \emph{consistency
condition},
\begin{equation}
\bar{M}_{j}^{i}=\bar{M}_{k}^{i}\bar{M}_{j}^{k}.\label{cons}%
\end{equation}
giving then rise to transitive matrices (see \cite{fa}). The most general
solution can be written as%
\begin{equation}
\frac{\bar{n}_{\alpha}^{i}}{\bar{n}_{\alpha}^{j}}=\bar{M}_{j}^{i}[S_{\alpha
}(\bar{M})]_{j}^{i},\label{nbar}%
\end{equation}
where $[S_{\alpha}(\bar{M})]_{j}^{i}$ are \emph{local} and
\emph{dimensionless} functions of the exchange rates, coming directly from the
preferences of the individual agents, and independent of the inventories. The
values $[S_{\alpha}(\bar{M})]_{j}^{i}$ must be\ then left invariant under the
transformation in Eq. (\ref{gauge}). Therefore, the solutions described by Eq.
(\ref{nbar}) are in the same equivalence class of the solutions of $\bar
{n}_{\alpha}^{i}=[S_{\alpha}(\bar{M}=1)]_{j}^{i}\bar{n}_{\alpha}^{j}$. It
follows that the space of equilibria at any point in time can be mapped to a
set $\{\bar{n}_{\alpha}^{1}:\alpha\in\mathcal{A}\}$ for a specific product,
say $1$. However, there is still a residual symmetry $\bar{n}_{\alpha}%
^{i}\mapsto\phi\bar{n}_{\alpha}^{i}$. We can use this symmetry to set
$n_{1}^{1}=1$. Therefore the landscapes of the equivalence classes of
equilibria at any point in time is the manifold $\left(  \mathbb{R}%
^{+}\right)  ^{n-1}$. When $n$ goes to infinity there is a one-to-one
correspondance between elements of this space and the distributions of wealth.
On the other hand the total non-equilibrium kinematic space is $(\mathbb{R}%
^{+})^{(n-1)\times p}$.

\section{Non-equilibrium dynamics}

Before proceeding, we need to make clear what we mean by \textquotedblleft
non-equilibrium". As we mentioned in the previous section, one can have
time-dependent equilibria. For example, suppose that agents have time
dependent preferences. We can then find a one-parameter family of
\emph{smooth} functions $\bar{n}_{\alpha}^{i}(t)$ and $\bar{M}_{j}^{i}(t)$,
constrained by $\frac{d}{dt}\sum_{\alpha}n_{\alpha}^{i}=0$, so that the
maximization conditions in Eq. (\ref{equilibrium}) are obeyed for any $t$. If
we regard $t$ as a time coordinate, such map would define a time-dependent
pure-exchange economy in equilibrium. Note that agents are still making
exchanges, but these are infinitesimal, \emph{i.e.}, $dn_{\alpha}^{i}\approx
dt$.

If the economy is out of equilibrium, the maximization conditions
(\ref{equilibrium}) are not obeyed. This means that there is an excess demand
or supply of products. Agents must then barter between each other in order to
find an equilibrium. Moreover, the amount that they will trade will be
\emph{finite}. In what follows we derive a set of equations to describe the
dynamics of the economy out of equilibrium. This is done under a minimal set
of assumptions which we shall discuss next.

Let $\bar{n}_{\alpha}^{i}(t)$ be a one-parameter family of equilibrium states
as discussed above. We can always decompose the agents' inventories as
$n_{\alpha}^{i}=\bar{n}_{\alpha}^{i}+\delta n_{\alpha}^{i}$, where $\delta
n_{\alpha}^{i}$ is a finite deviation from the particular equilibrium
trajectory $\bar{n}_{\alpha}^{i}(t)$. Let $\Delta n_{\alpha\beta}^{ij}$, be
the \emph{finite} amount of product $i$ that agent $\alpha$ gets (or gives) in
a trade with product $j$ and agent $\beta$ out of equilibrium. We assume that
$\Delta n_{\alpha\beta}^{ij}\approx\mathcal{O}(\delta n_{\alpha}^{i})$,
\emph{i.e.}, the (finite) corrections to the inventory are of the order of the
deviation from equilibrium. After a trade with agent $\beta$, agent $\alpha$
updates its inventory as $\delta n_{\alpha}^{i}\mapsto\delta n_{\alpha}^{i}+$
$\Delta n_{\alpha\beta}^{ij}$, $\delta n_{\alpha}^{j}\mapsto\delta n_{\alpha
}^{j}+\Delta n_{\alpha\beta}^{ji}$, $\delta n_{\alpha}^{k}\mapsto\delta
n_{\alpha}^{k}$, for $k\neq i,j$\text{. }By product conservation, we must have
$\Delta n_{\alpha\beta}^{ij}=-\Delta n_{\beta\alpha}^{ij}$ and trivially
$\Delta n_{\alpha\beta}^{ii}=0$. Finally, under the transformation given by Eq. (\ref{gauge}),
we must have $\Delta n_{\alpha\beta}^{ij}\mapsto\phi^{i}\Delta n_{\alpha\beta
}^{ij}$. By definition, $\Delta n_{\alpha\beta}^{ij}$ is the amount of product
$i,j$ that agents $\alpha$ and $\beta$ must exchange in order to be in
equilibrium with each other. The precise form of $\Delta n_{\alpha\beta}^{ij}$
can be calculated near equilibrium with an expansion in terms of the
fluctuations $\delta n$. Such an expansion is called \emph{perturbation
theory}. For now, we will work with a general $\Delta n_{\alpha\beta}^{ij}$
which obeys the basic properties given above.

In order to properly define a non-equilibrium economic theory, one must deal
with the question of how agents interacts. We will do this in a probabilistic
way. In this setting, time is continuous and when we take an infinitesimal
time interval $[t,t+dt]$, we assume that $dt$ is so small that any agent can
make \emph{at most} one barter process. Let $\sigma_{\alpha\beta}^{ij}(t)$ be
the probability per unit time that agent $\alpha$ will encounter agent $\beta$
and make a barter round involving product $i$ and $j$. The following facts are
evident: \emph{(i) }$\sigma_{\alpha\beta}^{ij}\geq0$ ; \emph{(ii) }%
$\sigma_{\alpha\beta}^{ij}=\sigma_{\beta\alpha}^{ij}$; \emph{(iii) }%
$\sigma_{\alpha\beta}^{ij}=\sigma_{\alpha\beta}^{ji}$; \emph{(iv) }%
$\sigma_{\alpha\beta}^{ii}=0$; \emph{(v) }$\sigma_{\alpha\alpha}^{ij}=0$.

If $\sigma_{\alpha\beta}^{ij}>0$ for every $\alpha,\beta\in\mathcal{A}$ then
the agents can be seen as located on the nodes of a complete network on $m$
nodes, \emph{i.e.}, a network in which every two nodes are connected by a
link. On the other hand, we may define $\sigma_{\alpha\beta}^{ij}$ based
directly on the structure of a chosen network. The network establishes a
constraing in the interaction between agents. Namely, given a network on $m$
nodes, we can associate the agents to the nodes and prescribe that
$\sigma_{\alpha\beta}^{ij}>0$ only if there is a link between $\alpha$ and
$\beta$.

Given an inventory at time $t$, $n_{\alpha}^{i}(t)$, the expected value at
time $t+dt$ of the non-equilibrium fluctuations, provided the information at
time $t$, is
\begin{align}
E_{t}[\delta n_{\alpha}^{i}(t+dt)] &  =\left(  1-dt\sum_{j,\beta}%
\sigma_{\alpha\beta}^{ij}(t)\right)  \delta n_{\alpha}^{i}(t)\label{nnew}\\
&  +dt\sum_{j,\beta}\sigma_{\alpha\beta}^{ij}(t)[\delta n_{\alpha}%
^{i}(t)+\Delta n_{\alpha\beta}^{ij}(t)].\nonumber
\end{align}
The first term comes from the probability of no interactions; the second term
is the contribution from the trading interactions. Taking an expectation value
of Eq. (\ref{nnew}) at time $t$ on both sides, we obtain an expression for the
unconditional expectations, which we denote by $\left\langle \cdot
\right\rangle $:%
\begin{equation}
\frac{d}{dt}\left\langle \delta n_{\alpha}^{i}(t)\right\rangle =\sum_{j,\beta
}\sigma_{\alpha\beta}^{ij}\langle\Delta n_{\alpha\beta}^{ij}(t)\rangle
.\label{nkin}%
\end{equation}
We can always write the fluctuations in terms of a scale invariant variable
$x_{\alpha}^{i}$, \emph{i.e.}, $\delta n_{\alpha}^{i}\equiv\bar{n}_{\alpha
}^{i}x_{\alpha}^{i}$. The evolution equations for the scale invariant
perturbations take the form%
\begin{equation}
\left(\frac{d}{dt}+\frac{1}{\bar{n}_{\alpha}^{i}}\frac{d\bar{n}_{\alpha}^{i}}%
{dt}\right)\left\langle x_{\alpha}^{i}(t)\right\rangle =\frac{1}{\bar{n}_{\alpha
}^{i}}\sum_{j,\beta}\sigma_{\alpha\beta}^{ij}\langle\Delta n_{\alpha\beta
}^{ij}(t)\rangle.\label{dxi}%
\end{equation}

In order to study perturbations in more detail, we need an explicit expression
for the solution of the bartering problem $\Delta n_{\alpha\beta}^{ij}$, and
for the probabilities $\sigma_{\alpha\beta}^{ij}$. It turns out that one can
find an explicit expression for $\Delta n_{\alpha\beta}^{ij}$, when agents are
assumed to be in a near-equilibrium state. This was given in \cite{v}. We
quote the result here:
\begin{equation}
\Delta n_{\alpha\beta}^{ij}\approx L\left(  \partial_{i}\delta\Omega
\partial_{j}\Omega-\partial_{i}\Omega\partial_{j}\delta\Omega\right)
\label{dni1}%
\end{equation}
and%
\begin{equation}
\Delta n_{\alpha\beta}^{ji}=-\left(  \partial_{i}\Omega/\partial_{j}%
\Omega\right)  \Delta n_{\alpha\beta}^{ij}; \label{dni2}%
\end{equation}%
\[
L=-\left(  \partial_{j}\Omega\right)  [\partial_{i}^{2}\Omega(\partial_{j}%
^{2}\Omega)^{2}+\partial_{j}^{2}\Omega(\partial_{i}^{2}\Omega)^{2}%
-2\partial_{i}\Omega\partial_{j}\Omega\partial_{i}\partial_{j}\Omega]^{-1},
\]
$\Omega:=\Omega_{\alpha}+\Omega_{\beta}$, and $\delta\Omega:=\Omega_{\alpha
}-\Omega_{\beta}$. At equilibrium, we know that $\bar
{M}_{j}^{i}=\partial_{i}\Omega_{\alpha}/\partial_{j}\Omega_{\alpha}$.
Therefore,
\begin{align*}
\Delta n_{\alpha\beta}^{ij}  &  =L\left(  \partial_{i}\delta\Omega\partial
_{j}\Omega-\partial_{i}\Omega\partial_{j}\delta\Omega\right) \\
&  =L(\partial_{i}\delta\Omega\partial_{j}\Omega(\bar{M}_{i}^{j}\bar{M}%
_{j}^{i}-1))\nonumber\\
&=0,
\end{align*}
as expected.

\section{A universality theorem}

In Section II, we showed that the only relevant scale invariant quantity in equilibrium
is the wealth distribution. Here we will show that under certain more
restrictive assumptions about the index of satisfaction, the final wealth
distribution can be found in terms of the initial conditions of the
perturbations, and it is \emph{independent on the details of the
non-equilibrium dynamics}. We show that this only happens if the following
conditions are met: in any of the possible equilibria, the index of
satisfaction is \emph{(i) }time-independent, \emph{i.e.}, $\left.
\partial_{t}\Omega_{\alpha}\right\vert _{\bar{n}}=0$; \emph{(ii) }the same for
all agents. These are necessary conditions for isolating the effects of the
network that determines interactions. However, we leave open the possibility
that, while out of equilibrium, the index of satisfaction might be time
dependent and agents might have different preferences. In fact, in Section V we will give a particular example of this case.

The key to our result, can be traced to the fact that under the conditions
given above, there are extra conserved quantities arising from Eq.
(\ref{nkin}). Under \emph{(i)} and \emph{(ii)}, the solution to the
equilibrium equation must have the form,
\begin{equation}
\frac{\bar{n}_{\alpha}^{i}}{\bar{n}_{\alpha}^{j}}=\bar{M}_{j}^{i}\frac{\nu
^{i}(\bar{M})}{\nu^{j}(\bar{M})}\label{equil}%
\end{equation}
for any equilibrium state. Here we have written $(S_{\alpha})_{j}^{i}%
:=S_{j}^{i}\equiv\nu^{i}/\nu^{j}$, since \emph{(ii)}. The fact that we can
decompose $S_{j}^{i}$ as above, follows from the consistency relations that
$S_{j}^{i}$ must obey (see Eqs. (\ref{cons}) and \ref{nbar}). The functions
$\nu^{i}$ are assumed to be time-independent from \emph{(i)}, therefore the
possible equilibria will also be time independent. We have already seen that
product number is conserved in the barter process. Thus, from Eq. (\ref{nkin})
we have $\frac{d}{dt}\sum_{\alpha}\bar{n}_{\alpha}^{i}x_{\alpha}^{i}=0$.
However, there are more subtle conservation laws not evident from Eq.
(\ref{nkin}), but arising in the linearized approximation. Under this
approximation we can write
\begin{equation}
\Delta n_{\alpha\beta}^{ji}\approx-(\bar{M}_{i}^{j}+\mathcal{O}(\delta))\Delta
n_{\alpha\beta}^{ij}.\label{linear}%
\end{equation}
That is, agents are trading at approximately the same exchange rate as in
equilibrium. Using Eqs. (\ref{equil}) in (\ref{linear}) one can easily show
that
\begin{equation}
\frac{\nu^{i}(\bar{M})\Delta n_{\alpha\beta}^{ij}}{\bar{n}_{\alpha}^{i}%
}\approx-\frac{\nu^{j}(\bar{M})\Delta n_{\alpha\beta}^{ji}}{\bar{n}_{\alpha
}^{j}}.\label{linear1}%
\end{equation}
By Eq. (\ref{linear1}) in Eq. (\ref{nkin}) it follows that 
\begin{equation}\frac{d}{dt}%
\sum_{i}\nu^{i}x_{\alpha}^{i}=0\;.
\end{equation}

The next step is to find the asymptotic prices at $t\rightarrow\infty$. Of
course, we need to assume that the system reaches some other equilibrium
state. Hence,
\begin{equation}
M_{j}^{i}(\infty)\frac{\nu^{i}(M(\infty))}{\nu^{j}(M(\infty))}=\frac
{n_{\alpha}^{i}(\infty)}{n_{\alpha}^{j}(\infty)}=\frac{\bar{n}_{\alpha}%
^{i}(1+x_{\alpha}^{i}(\infty))}{\bar{n}_{\alpha}^{j}(1+x_{\alpha}^{j}%
(\infty))}. \label{Meq}%
\end{equation}
Note that in writing Eq. (\ref{Meq}) we have assumed that the form of $\nu
^{i}$ as a function of prices is the same at $t\rightarrow\infty$ that at
$t=0$. This follows from $(i)$. Summing over $\mathcal{A}$, under the
linearized approximation, we get
\begin{equation}
M_{j}^{i}(\infty)\frac{\nu^{i}(M(\infty))}{\nu^{j}(M(\infty))}\approx\bar
{M}_{j}^{i}\frac{\nu^{i}(\bar{M})}{\nu^{j}(\bar{M})}\left(  1+\frac{J^{i}%
}{\bar{n}^{i}}-\frac{J^{j}}{\bar{n}^{j}}\right)  , \label{Mfinal}%
\end{equation}
where $\bar{n}^{i}=\frac{1}{m}\sum_{\alpha}\bar{n}_{\alpha}^{i}$ and
$J^{i}=\frac{1}{m}\sum_{\alpha}\bar{n}_{\alpha}^{i}x_{\alpha}^{i}$. Note that
$J^{i}$ are conserved quantities, and hence are given in terms of the initial
conditions of the perturbations. One can then use Eq. (\ref{Mfinal}) to solve
for the final prices in terms of the equilibrium state we are expanding
around, and the initial conditions for the perturbations. We are now in the
position to deal with the final wealth distribution. In units of product $i$,
we have
\begin{align*}
W_{\alpha}^{i}(\infty)  &  =\sum_{k}M_{k}^{i}(\infty)n_{\alpha}^{k}(\infty)\\
&  \approx\sum_{k}M_{k}^{i}(\infty)\bar{n}_{\alpha}^{k}+\sum_{k}\bar{M}%
_{k}^{i}\bar{n}_{\alpha}^{k}x_{\alpha}^{k}(\infty)\\
&  =\sum_{k}M_{k}^{i}(\infty)\bar{n}_{\alpha}^{k}+\frac{p\bar{n}_{\alpha}^{i}%
}{\nu^{i}(\bar{M})}X_{\alpha},
\end{align*}
where $X_{\alpha}:=\frac{1}{p}\sum_{i=1}^{p}\nu^{i}x_{\alpha}^{i}$. We note
that this is also a conserved quantity, and so it is given by its initial
value. Since $M_{k}^{i}(\infty)$ can be determined from Eq. (\ref{Mfinal}) in
terms of the initial perturbations, we have the followin result:

Consider a pure exchange economy, with the space for
all equivalence classes of equilibria determined by the scale transformation in Eq. (\ref{gauge}). Let us assume that the index of satisfaction at equilibrium is
\emph{(i) }time-independent and \emph{(ii) }the same for all agents. Moreover,
let us assume that given an initial non-equilibrium linearized fluctuation,
the system will go back to some equilibrium state at time $t\rightarrow\infty
$. Thus, the wealth distribution is completely determined in terms of the
initial conditions of the perturbation and it is independent on the
non-equilibrium dynamics. In particular, it is independent of the
probabilities and the network of interaction between agents.

As a special case, consider an homogeneous equilibrium state $\bar{n}_{\alpha
}^{i}\equiv\bar{n}^{i}$. In this case, all agents have the same wealth,
$\bar{W}^{i}$, say. For simplicity, assume that $\nu^{i}=1$. Then, given an
initial perturbation $\delta n_{\alpha}^{i}\equiv\bar{n}^{i}x_{\alpha}^{i}$,
the final wealth of the economy is given by,
\begin{align}
W_{\alpha}^{i}(\infty)  &  \approx\bar{W}^{i}\left(  1+\frac{J^{i}}{\bar
{n}^{i}}-\frac{J^{j}}{\bar{n}^{j}}+X_{\alpha}\right) \label{finalW}\\
&  =\bar{W}^{i}\left(  1+\frac{1}{m}\sum_{\alpha}[x_{\alpha}^{i}(0)-x_{\alpha
}^{j}(0)]+\frac{1}{p}\sum_{i}x_{\alpha}^{i}(0)\right)  .\nonumber
\end{align}
In particular, we see that since $X_{\alpha}$ is conserved, the final
equilibrium state of the economy will generically be non-homogeneous. However,
note that if $X_{\alpha}=X_{\beta}$ for every $\alpha$ and $\beta$, the final
wealth distribution will again be homogeneous, and thus related by a a scale
transformation to $\bar{W}^{i}$.

\section{Examples}

\subsection{Homogeneous static preferences}

We discuss here a particular example to clarify the role of the probabilities
$\sigma_{\alpha\beta}^{ij}$, the network of interaction, and the stability of
the non-equilibrium perturbations. Agents are associated to the nodes of a
fixed network. Each agent can interact with exactly $d$ other fixed agents,
\emph{i.e.}, the network is modeled by a $d$-regular graph. Let $A$ be the
adjacency matrix of the network: $A_{\alpha\beta}=1$ if $\sigma_{\alpha\beta
}^{ij}>0$ for at least two products $i$ and $j$; $A_{\alpha\beta}=0$,
otherwise. When the probability of trading two specific products is uniform,
we have $\sigma_{\alpha\beta}^{ij}=A_{\alpha\beta}/d(p-1)$. We study in the
following case of homogeneous time-independent index of satisfaction:%
\begin{equation}
\Omega_{\alpha}=\sum_{i}\log n_{\alpha}^{i}.\label{util1}%
\end{equation}
This index trivially satisfies all the previous assumptions. Therefore, the
final wealth distribution will be independent on the dynamics. In this simple example, one can obtain an analytical solution to the bartering problem:  $\Delta n_{\alpha
\beta}^{i}=(n_{\alpha}^{j}n_{\beta
}^{i}-n_{\alpha}^{i}n_{\beta}^{j})/(2(n_{\alpha}^{j}+n_{\beta}^{j}))$. Next, we note that the equilibrium equations
(\ref{equilibrium}) for this system reduce to $\bar{M}_{j}^{i}=\bar{n}%
_{\alpha}^{i}/\bar{n}_{\alpha}^{j}$.  It is then easy to see from
Eqs. (\ref{dni1}) and (\ref{dni2}) that ${\Delta n_{\alpha\beta}^{ij}%
}/{\bar{n}_{\alpha}^{i}}\approx\frac{1}{4}(x_{\alpha}^{j}-x_{\alpha}%
^{i}-x_{\beta}^{j}+x_{\beta}^{i})$. Therefore the equation for the scale
invariant perturbations becomes
\begin{equation}
\frac{d}{dt}\langle x_{\alpha}^{i}\rangle=\frac{1}{4d(p-1)}\sum_{\beta
,j}A_{\alpha\beta}(\langle x_{\alpha}^{j}\rangle-\langle x_{\alpha}^{i}%
\rangle-\langle x_{\beta}^{j}\rangle+\langle x_{\beta}^{i}\rangle
){.}\label{dxi1}%
\end{equation}
These equations can be seen as the direct analog of the wealth dynamic equations proposed in \cite{bm}. Therefore, our perturbation theory can be seen as providing a microeconomic foundation to the results of \cite{bm}. However, we do not include a stochastic source term, which would keep the system out of equilibrium. The presence of such a term can be directly linked to the fat-tailed wealth distribution obtained in \cite{bm}. 

\begin{figure}
[h]
\begin{center}
\includegraphics[
height=2.0237in,
width=3.275in
]%
{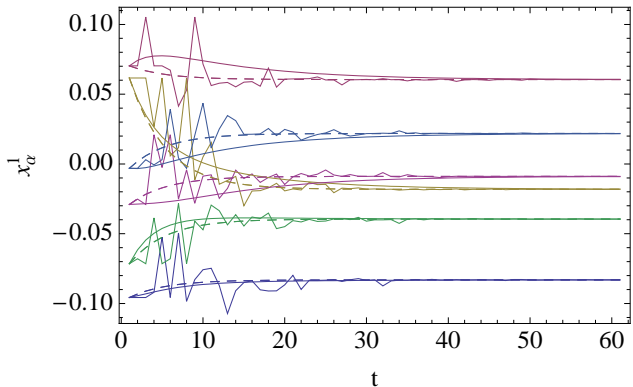}%
\caption{ The dashed and solid curves represent the predicted
fluctuations in terms of the scale invariant variable $x_{\alpha}^{1}$ as a
funtion of time $t$ for agents on the complete graph and the cycle graph on
$6$ vertices, respectively. Since the eigenvalue gap for the complete graph is
larger than the one for the cycle, the rate of converge is higher for the
complete graph. The spiked curves are obtained by a numerical simulation. This
is consistent with the mean field theory.}%
\label{fig1}%
\end{center}
\end{figure}

We can readily see from Eq. (\ref{dxi1}) the conserved quantities found in the last section:
$\frac{d}{dt}\sum_{\alpha}x_{\alpha}^{i}=0$ and $\frac{d}{dt}\sum_{i}%
x_{\alpha}^{i}=0$. Hence, the final wealth distribution is given by Eq.
(\ref{finalW}). We are interested in showing that there are no instabilities,
and so the economy indeed equilibrates. The role of the network is to
determine the convergence rate towards equilibrium, in evident analogy with
the notion of mixing time for random walks on graphs (see \cite{af}). Let us
write Eq. (\ref{dxi1}) in vector notation as, $
\frac{d}{dt}\vec{x}=\Gamma
\cdot\vec{x}$, 
where $\vec{x}=\left(  x_{1}^{1},\ldots,x_{N}^{1},x_{1}%
^{2},\ldots x_{N}^{2},\ldots\right)  ^{T}$ and
\[
\Gamma=\frac{1}{4(p-1)}(J_{p}-pI_{p})\otimes(I_{m}-T),
\]
being $I_{n}$ the $m\times m$ identity matrix, $J_{p}$ the $p\times p$ all-one
matrix, and $T:=A_{\alpha\beta}/d$. The eigenvalues of $\Gamma$ are
$\gamma_{\alpha}^{i}=\frac{1}{4(p-1)}(\mu^{i}-p)(1-\lambda_{\alpha})$, where
$\mu^{i}\in\{0,p\}$ and $\lambda_{\alpha}$ are the eigenvalues of $J_{p}$ and
$T$ respectively. It follows that there are $m$ steady states associated with
$\gamma_{\alpha}^{1}=0$. These correspond to the conserved quantities
$\sum_{i}x_{\alpha}^{i}$. The remaining eigenvalues are $\gamma_{\alpha}%
^{i}=-\frac{p}{4(p-1)}(1-\lambda_{\alpha})$. Because $T$ is doubly stochastic,
$\lambda_{\alpha}\in\lbrack-1,1]$ and $\lambda_{1}=1$. So, $\gamma_{\alpha
}^{i}\leq0$ and there is no instability. The rate of convergence is determined
by the \emph{eigenvalue gap} $\left\vert \lambda_{1}-\lambda_{2}\right\vert $
of the matrix $T$. The network does not affect the wealth distribution but it
gives the rate of convergence towards equilibrium. As expected, the larger the
eigenvalue gap, the faster is the convergence. Graphs with good expansion properties are then associated with fast relaxation. Fig. \ref{fig1} gives an example
for the complete graph and the cycle graph on six vertices, using the same initial conditions. The numerical
simulation is consistent with the mean field theory. The simulations are obtained taking Eq. (\ref{dxi1}) in discrete time. Moreover, we replace the matrix $A_{\alpha\beta}/d$ by a {\it random} matrix which chooses one pair of agents to interact in every time step. The pair is chosen from a uniform random distribution.

%
%TCIMACRO{\FRAME{fhFU}{3.275in}{2.0237in}{0pt}{\Qcb{{}(Color online) The dashed
%and solid curves represent the predicted fluctuations in terms of the scale
%invariant variable $x_{\alpha}^{i}$ as a funtion of time $t$ for agents on the
%complete graph and the cycle graph on $6$ vertices, respectively. Since the
%eigenvalue gap for the complete graph is larger than the one for the cycle,
%the rate of converge is higher for the complete graph. The spiked curves are
%obtained by a numerical simulation. This is consistent with the mean field
%theory.}}{\Qlb{fig1}}{plot1.eps}{\special{ language "Scientific Word";
%type "GRAPHIC";  maintain-aspect-ratio TRUE;  display "PICT";
%valid_file "F";  width 3.275in;  height 2.0237in;  depth 0pt;
%original-width 4.1632in;  original-height 2.559in;  cropleft "0";
%filename 'plot1.eps';file-properties "XNPEU";}} }%
%BeginExpansion
%EndExpansion
%TCIMACRO{\FRAME{fbhFU}{3.2742in}{2.0237in}{0pt}{\Qcb{(Color online) this is
%the caption this is the caption this is the caption this is the caption this
%is the caption this is the caption this is the caption this is the caption
%this is the caption this is the caption this is the caption this is the
%caption this is the caption this is the caption this is the caption.}%
%}{\Qlb{fig2}}{plot3.eps}{\special{ language "Scientific Word";
%type "GRAPHIC";  maintain-aspect-ratio TRUE;  display "PICT";
%valid_file "F";  width 3.2742in;  height 2.0237in;  depth 0pt;
%original-width 4.1632in;  original-height 2.559in;  cropleft "0";
%filename 'plot3.eps';file-properties "XNPEU";}} }%
%BeginExpansion
\begin{figure}
[hbh]
\begin{center}
\includegraphics[
height=2.0237in,
width=3.2742in
]%
{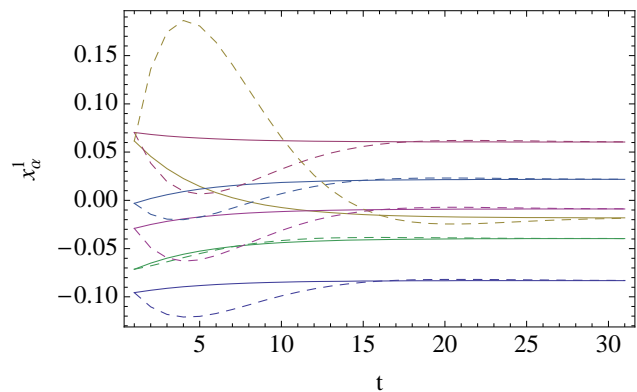}%
\caption{ Evolution of the scale invariant variable $x_{\alpha}^{1}$ as a
funtion of time $t$ for agents on the complete graph with
$6$ vertices. The solid lines are for the example with homogeneous unchanging preferences, Eq. (\ref{dxi1}). The dashed lines are for the case of heterogeneous dynamic preferences, Eqs. (\ref{dxi2}) and (\ref{dy}).  We have taken $\nu = 0.9$ for the heterogeneous case.}%
\label{fig2}%
\end{center}
\end{figure}
%EndExpansion

\subsection{Heterogeneous dynamic preferences}

It is worth considering an example with a slightly more complicated index of
satisfaction,%
\begin{equation}
\Omega_{\alpha}=\log\sum_{j}[(M_{\alpha})_{j}^{i}n_{\alpha}^{j}]^{\nu
},\label{util2}%
\end{equation}
where $0<\nu<1$ is a fixed but arbitary constant. The matrix $M_{\alpha}$
represents a particular preference of the agent. It can be interpreted as an
opinion about the exchange rates. Different agents might have different
opinions. We assume that $M_{\alpha}$ obeys the usual consistency conditions.
Therefore, the choice of index $i$ in Eq. (\ref{util2}) is arbitrary. We will
build a model where agents can learn the ``market" exchange
rates when they barter with other agents. This means, they will ultimately
converge to some equilibrium prices $\bar{M}$. Therefore,\emph{ for any of
such equilibria}, the preferences will be of the form:%
\[
\bar{\Omega}_{\alpha}=\log\sum_{j}[\bar{M}_{j}^{i}\bar{n}_{\alpha}^{j}]^{\nu}.
\]
One can easily show, from Eqs. (\ref{equilibrium}), that thee space of
equilibria of this model is the same as in the previous one: $\bar{M}_{j}%
^{i}=\bar{n}_{\alpha}^{i}/\bar{n}_{\alpha}^{j}$. Here we will expand, as in
the previous case, around an homogeneous economy with $\bar{n}_{\alpha}%
^{i}=\bar{n}^{i}$. Note that even though in this model agents have
time-dependent preferences out of equilibrium, they all converge to
homogeneous preferences in equilibrium. This means that this model obeys the
simplifying assumptions of Section II, and hence the final wealth distribution
is independent of the non-equilibrium dynamics.

Now we need to provide a model of how the agents learn the new prices. We
assume that, when two agents $\alpha$ and $\beta$ find each other and make a
trade, they update their respective matrices $M_{\alpha}$ and $M_{\beta}$ as%
\begin{equation}
(M_{\alpha})_{j}^{i}=(M_{\beta})_{j}^{i}=-\frac{\Delta n_{\alpha\beta}^{ij}%
}{\Delta n_{\alpha\beta}^{ji}}\approx\frac{\partial_{j}(\Omega_{\alpha}%
+\Omega_{\beta})}{\partial_{i}(\Omega_{\alpha}+\Omega_{\beta})}%
,\label{Mupdate}%
\end{equation}
\ for all $i,j\in\mathcal{P}$. We have used Eqs. (\ref{dni1}) and
(\ref{dni2}). This way of updating prices makes sure that the exchange rates
$M_{\alpha}$ always obey the consistency condition (see Eq.( \ref{cons})).
Moreover, one could interpret such updating as an exchange of information
between both agents: even though agents traded only two products, they
\textquotedblleft found out\textquotedblright\ about each other's prices. The
expectation of $\alpha$'s internal matrix $M_{\alpha}$ at the next time step is
\begin{align}
E_{t}[(M_{\alpha})_{j}^{i}(t+dt)]  & =\left(  1-dt\sum_{\beta}T_{\alpha\beta
}(t)\right)  (M_{\alpha})_{j}^{i}(t)\label{EM}\\
& +dt\sum_{\beta}T_{\alpha\beta}(t)\frac{\partial_{j}(\Omega_{\alpha}%
+\Omega_{\beta})}{\partial_{i}(\Omega_{\alpha}+\Omega_{\beta})}.\nonumber
\end{align}
We have used the fact that the total probability per unit time of an encounter
between agents $\alpha$ and $\beta$ is given by $\sum_{j}\sigma_{\alpha\beta
}^{ij}=T_{\alpha\beta}$. Taking expectations on both sides of Eq. (\ref{EM})
we get,
\begin{equation}
\frac{d}{dt}\langle(M_{\alpha})_{j}^{i}(t)\rangle=-\langle(M_{\alpha})_{j}%
^{i}(t)\rangle+\sum_{\beta}T_{\alpha\beta}(t)\langle\frac{\partial_{j}(\Omega_{\alpha}%
+\Omega_{\beta})}{\partial_{i}(\Omega_{\alpha}+\Omega_{\beta})}\rangle.\label{EMFinal}%
\end{equation}
This can be seen as the evolution equation for the agent's preferences.

We are now ready to work out the evolution equations for the scale invariant
perturbations. Due to the consistency condition, we can always consider only
the $(M_{\alpha})_{i}^{1}$ components of the prices. We can then define the
following scale invariant variables for the price perturbations: 
\begin{equation}y_{\alpha
}^{i}:=\frac{(M_{\alpha})_{i}^{1}(t)-\bar{M}_{i}^{1}}{\bar{M}_{i}^{1}}\;.
\end{equation}
 One
finds that for the utility in Eq. (\ref{util2}), and for the homogeneous
background, the amount of products traded is%
\begin{eqnarray}
\label{DeltaN}
\frac{\Delta n_{\alpha\beta}^{ij}}{\bar{n}^{i}}  & =&\frac{1}{4}[x_{\alpha}%
^{j}-x_{\beta}^{j}-x_{\alpha}^{i}+x_{\beta}^{i}\label{dnij}\nonumber \\
&& +\frac{\nu}{1-\nu}(y_{\alpha}^{i}-y_{\beta}^{i}-y_{\alpha}^{j}+y_{\beta}%
^{j})], 
\end{eqnarray}
where it is understood that $y_{\alpha}^{1}\equiv0$. By plugging Eq. (\ref{DeltaN}) in
Eq. (\ref{dxi}), we get the equation for the perturbation in inventory:%
\begin{eqnarray}
\label{dxi2}
\frac{d}{dt}\langle x_{\alpha}^{i}\rangle & =&\frac{1}{4(p-1)}\sum
_{\alpha,\beta}T_{\alpha\beta}[\langle x_{\alpha}^{i}\rangle-\langle x_{\beta
}^{j}\rangle-\langle x_{\alpha}^{i}\rangle+\langle x_{\beta}^{i}%
\rangle\nonumber \\
&& +\frac{\nu}{1-\nu}(\langle y_{\alpha}^{i}\rangle-\langle y_{\beta}%
^{j}\rangle-\langle y_{\alpha}^{i}\rangle+\langle y_{\beta}^{i}\rangle
)].
\end{eqnarray}
Next we derive the equation for the perturbations in the internal prices of
the agents. In order to do this, we need the following result:
\begin{eqnarray}
\frac{\partial_{i}\Omega}{\partial_{j}\Omega}  & \approx&1+\frac{1}{2}%
(1-\nu)(x_{\alpha}^{j}+x_{\beta}^{j}-x_{\alpha}^{i}-x_{\beta}^{i}%
)\label{dOmega}\nonumber \\
&& +\frac{1}{2}\nu(y_{\alpha}^{i}+y_{\beta}^{i}-y_{\alpha}^{j}-y_{\beta}%
^{j}),
\end{eqnarray}
where we have set $\bar{M}_{j}^{i}=1$ without any loss of generality. We can now use Eq. (\ref{dOmega}) in Eq.
(\ref{EMFinal}) to get an expression for the price perturbations:%
\begin{align}
\frac{d}{dt}{\langle}y_{\alpha}^{i}{\rangle}  & {=}-\langle y_{\alpha}%
^{i}\rangle+\frac{1}{2}\sum_{\beta}T_{\alpha\beta}[(1-\nu)(\langle x_{\alpha
}^{1}\rangle+\langle x_{\beta}^{1}\rangle\label{dy}\\
& -\langle x_{\alpha}^{i}\rangle-\langle x_{\beta}^{i}\rangle)+\nu\left(
\langle y_{\alpha}^{i}\rangle+\langle y_{\beta}^{i}\rangle\right) ]\;,\nonumber
\end{align}
{w}here, again, $y_{\alpha}^{1}\equiv0$. We can now analyze Eqs. (\ref{dxi2})
and (\ref{dy}) in order to prove the stability of the system. It is useful to
diagonalize the matrix $T$, as in the previous example. Recall that the price
perturbation $y_{\alpha}^{1}\equiv0$, since this is the perturbation of
$(M_{\alpha})_{1}^{1}=1$. However, we can just include $y_{\alpha}^{1}$ as a
spurious variable which will be conserved, \emph{i.e.}, $\frac{d\langle
y_{\alpha}^{1}\rangle}{dt}=0$. One can then show that Eqs. (\ref{dxi2}) and
(\ref{dy}) can be written in matrix form as $\frac{d}{dt}\vec{X}=\Gamma
\cdot\vec{X}$, where $\vec{X}=(x_{1}^{1},\ldots x_{n}^{1},x_{1}^{2}%
,\ldots,x_{n}^{p},y_{1}^{1},\ldots,y_{n}^{1},y_{1}^{2},\ldots,y_{n}^{p})^{T}$
and
\[
\Gamma=\left(
\begin{array}
[c]{cc}%
\frac{1}{4(p-1)}a_{1} & -\frac{\nu}{4(p-1)(1-\nu)}a_{1}\\
a_{2} & a_{3}%
\end{array}
\right)  .
\]
Here $R_{p}$ is the $p\times p$ matrix with entries $(R_{p})_{ij}=\delta
_{j,1}$. Moreover,
 \begin{eqnarray*}
 a_{1}&=&(J_{p}-pI_{p})\otimes(I_{n}-T)\;,\\
 a_{2}&=&(1-\nu)(R_{p}%
-I_{p})/2\otimes(I_{n}+T)\;, \\
a_{3}&=&(R_{p}-I_{p})\otimes(I_{n}-(I_{n}%
+T)\nu/2)\;.\end{eqnarray*}
 When we take two products ($p=2$), the two non-zero eigenvalues of
$\Gamma$ are
\begin{eqnarray}
\label{eigenvalues} 
\gamma_{\pm}&=&\frac{1}{4}\left[-3+\nu\lambda+\lambda+\nu\right.\nonumber \\
&&\left. \pm
\sqrt{(\nu\lambda+\lambda+\nu-3)^{2}+8(\lambda-1)}\right]\;,
\end{eqnarray} with $\nu
\in(0,1)$ and $\lambda\in\lbrack-1,1]$. It is straightforward to see that {\it the real part} of both
eigenvalues is negative or zero. If we fix $\nu$ then the eigenvalue gap
still determines the convergence rate as in the case of homogeneous
preferences. 

In Fig. \ref{fig2}, we compare the dynamics of this model with the simpler one of the previous section. We take the same initial conditions for the scale invariant perturbations, and fix the network to be the complete graph. We see that, as expected, the final state is the same for both models. However, for the changing preferences model, there are large oscillations due to the fact that the eigenvalues given in Eq. (\ref{eigenvalues}) are complex. Moreover, in the limit $\nu \rightarrow 0$ (wealth maximizers), these oscillations dominate and the system never equilibrates.

\section{Conclusions}
In this paper we have introduced a theory of linearized perturbations around a pure exchange economy. Our formalism is given in terms of a general index of satisfaction, and the probabilities of agents interacting on a network. We have shown that, if agents have static preferences, which
are also homogeneous in any of the steady states, the final wealth
distribution is independent of the dynamics of the non-equilibrium theory. In
particular, it is completely determined in terms of the initial conditions, and it is independent of the
probability and the network of interaction between agents. We have shown that the
main effect of the network is to determine the relaxation time to equilibrium. We gave two examples where the relaxation times can be computed analytically. Moreover, we showed the agreement between the mean field theory technique and numerical simulations. 

This work can be extended in a number of directions. First, it would be interesting to consider agents that have random changing preferences, or can speculate on exchange rates. Some simulations of this kind were conducted in \cite{v} in the context of a centralized market. These speculations can lead to stochastic terms analog to the ones proposed in \cite{bm}. Finally, we have only considered agents making barter exchanges at a fixed time. An important component of the economy is that agents are able to make exchanges at different times. That is, we have contingency claims. It would be interesting to set up a similar perturbation theory involving such claims, and study the stability of the system.

\bigskip

\noindent\emph{Acknowledgments. }The authors would like to thank Mike Brown,
Jim Herriot, Stuart Kauffman, Zoe-Vonna Palmrose, and Lee Smolin, for helpful
discussion. S. E. V. would like to thank Maes Bert for his kind hospitality at
the Department of Chemistry of the University of Antwerp, where part of this
work has been done. Research at Perimeter Institute for Theoretical Physics is
supported in part by the Government of Canada through NSERC and by the
Province of Ontario through MRI. Research at IQC is supported in part by
DTOARO, ORDCF, CFI, CIFAR, and MITACS.

\end{document}